\documentclass[10pt,twocolumn,letterpaper]{article}

\usepackage{iccv}              
\usepackage[accsupp]{axessibility}
\usepackage{amsmath}
\usepackage{amsthm}
\usepackage{colortbl}
\usepackage{multirow}
\usepackage{adjustbox}
\usepackage{geometry}
\usepackage{supertabular}
\usepackage{colortbl}
\usepackage{diagbox}
\definecolor{lavender}{rgb}{0.9, 0.9, 0.98}
\definecolor{iccvblue}{rgb}{0.21,0.49,0.74}
\definecolor{BoldDelta}{HTML}{287EB8}
\usepackage[pagebackref,breaklinks,colorlinks,allcolors=iccvblue]{hyperref}
\usepackage{tabularx}
\usepackage{booktabs}       
\usepackage{caption}
\def\confName{ICCV}
\def\confYear{2025}

\title{ Test-time Adaptation for Foundation Medical Segmentation Model without Parametric Updates}

\author{
Kecheng Chen$^{1}$,
Xinyu Luo$^{1}$ ,
Tiexin Qin$^{1}$ ,
Jie Liu$^{1}$ ,
Hui Liu$^{1}$, \\
Victor Ho Fun Lee$^{2}$ ,
Hong Yan$^{1}$ ,
Haoliang Li$^{1}$*\\
\vspace{2mm}
$^{1}$ City University of Hong Kong\quad
$^{2}$HKU
{\tt\small \{ck.ee@my.cityu.edu.hk\} \quad *:Corresponding author}
}

\begin{document}
\maketitle
\begin{abstract}
Foundation medical segmentation models, with MedSAM being the most popular, have achieved promising performance across organs and lesions. However, MedSAM still suffers from compromised performance on specific lesions with intricate structures and appearance, as well as bounding box prompt-induced perturbations.  Although current test-time adaptation (TTA) methods for medical image segmentation may tackle this issue, partial (e.g., batch normalization) or whole parametric updates restrict their effectiveness due to limited update signals or catastrophic forgetting in large models. Meanwhile, these approaches ignore the computational complexity during adaptation,  which is particularly significant for modern foundation models. To this end,  our theoretical analyses reveal that directly refining image embeddings is feasible to approach the same goal as parametric updates under the MedSAM architecture, which enables us to realize high computational efficiency and segmentation performance without the risk of catastrophic forgetting. Under this framework, we propose to encourage maximizing factorized conditional probabilities of the posterior prediction probability using a proposed distribution-approximated latent conditional random field loss combined with an entropy minimization loss. Experiments show that we achieve about 3\% Dice score improvements across three datasets while reducing computational complexity by over 7 times.
\end{abstract}    
\section{Introduction}
\label{sec:intro}
Recently, foundation segmentation models for medical images have attracted massive attention from the medical imaging community~\citep{zhang2023input}, as medical image segmentation plays an indispensable role in many downstream tasks (\textit{e.g.,} surgical navigation and treatment planning). Medical Segment Anything Model~\citep{ma2024segment} (MedSAM) is the most popular foundation segmentation model, which exhibits impressive universality across organs and lesions. Since MedSAM is pre-trained on large-scale medical image datasets, it typically enables the high-quality segmentation mask of the region of interest (RoI) in a zero-shot manner, which has contributed to extensive areas, \textit{e.g.,} semi/weak-supervised medical image segmentation~\citep{xu2024sam} or anomaly detection~\citep{maosemi} using provided segmentation masks as pseudo labels or RoIs.
\begin{table}[!t]
    \renewcommand\arraystretch{1.}
    \centering
  
    \resizebox{0.49\textwidth}{!}{%
    \setlength\tabcolsep{6.0pt}
    \scalebox{1.00}{
    \begin{tabular}{c|c|c c c|c}
        \toprule
        \multirow{2}{*}{Update Type} & \multirow{2}{*}{Method} & \multicolumn{3}{c|}{\textbf{Datasets}} & \multirow{2}{*}{\textbf{GFLOPs}$\downarrow$} \\ 
        \cline{3-5}
        & & OC & OD & SCGM & \\ 
        \hline 
        \rowcolor{gray!10}    &  Direct Inference & 87.77 & 80.57 & 56.67 & - \\ 
        \hline
        \multirow{4}{*}{Normalization} 
         & TENT~\citep{wang2020tent} (ICLR'21)& 86.24 & 78.08 & 54.34 & 743.5 \\ 
         & InTEnt~\citep{dong2024medical} (CVPR'24) &  86.95& 82.40 & 49.81 & 744.8 \\ 
         & GraTa~\citep{chen2025grata} (AAAI'25)& 88.57 & 80.41& 55.14  & 2975.6\\ 
         & PASS~\citep{zhang2024pass} (TMI'24)&84.05  & 71.81 & 54.80 &745.8 \\ \hline
        \multirow{2}{*}{Full} 
         & MEMO~\citep{zhangmemo} (NeurIPS'22)& 88.12 & 80.04 & 57.48 & 5646.5\\ 
         & CRF-SOD~\citep{veksler2023test} (CVPR'23)&  88.95& 79.79 & 56.81 & 1115.2 \\ \hline
        \rowcolor{gray!5} Latent &  \textit{Ours} & \textbf{90.84} & \textbf{84.19} & \textbf{59.49} &\textcolor{red}{\textbf{97.3}} \\ \bottomrule
    \end{tabular}
    }}
     \caption{\small{Average Dice coefficient ($\uparrow$) of different TTA methods on three datasets. Normalization-based methods are customized to BN layers, exhibiting marginal (even degraded) performance for LN-based foundation medical segmentation model. We report the computational complexity using GFLOPs (Giga Floating Point Operations), including forward and backward propagations. \textbf{We leverage latent updates to achieve $\textcolor{red}{\sim \textbf{7}\times}$ GFLOPs reduction compared with parametric update-based counterparts with performance gains.} }}
    \label{gflops}
    \vspace{-0.5cm}
    \end{table}

However, due to the extreme modality imbalance of medical images (where CT, MRI, and endoscopy images are dominant), MedSAM has compromised performance on less-represented modalities (\textit{e.g.}, eye fundus images)~\citep{ma2024segment}. It is also difficult for MedSAM to 
maneuver specific lesions with intricate structures in the segmentation area~\citep{ma2024segment}, \textit{e.g.,} optic cup segmentation suffers from many vessel-like branching structures. Although supervised fine-tuning on specific modalities and lesions with annotated masks seems effective in tackling these issues~\citep{wu2023medical}, the overwhelming computational consumption of fine-tuning and inherent data scarcity (or privacy) concerns regarding annotated medical images may impede this process. Moreover, some test-time perturbations are still encountered, even though a well-calibrated and fine-tuned model has been prepared. For instance, mild looseness or shrinkage perturbations for bounding box prompts are sufficient to cause performance degradation for MedSAM (see Tables \ref{tab:comparisons_fundus_OC} and \ref{tab:comparisons_fundus_OD}), which is clinically unfavorable as delineating a perfect prompt is usually labor-intensive and time-consuming for physicians.

To address the above-mentioned issues, test-time adaptation (TTA) for foundation medical segmentation models is a feasible solution that enables performance gains on a specific unlabeled target (test) domain using the pre-trained model only. Although recent TTA approaches for medical image segmentation achieve promising results, we argue that scaling them to foundation segmentation models is difficult. First, most methods 
utilize various self-supervised objectives (\textit{e.g.,} prediction consistency over different augmentations~\citep{chen2024gradient,dong2024medical,zhang2024pass}) to conduct batch normalization (BN)-centric fine-tuning, which may endure limited update signals with marginal adaptation as modern foundation models employ layer normalization (LN) without source-trained statistics like BN. 
Second, some approaches~\citep{wen2024denoising} update the whole model to explore more gains, but even minimal parameter adjustments in large models can precipitate catastrophic forgetting~\citep{ju-etal-2024-mitigating}. Third, most methods ignore computational complexity during the adaptation, which is particularly significant for foundation medical segmentation models\footnote{MedSAM obtains image embedding using an intricate ViT-based encoder (accounts for 98.5\% computations) and a lightweight decoder.}, since continuous forward and backward propagations on a large model are computationally intensive, especially on resource-constrained edge devices.  \textit{Overall, existing methods struggle with maintaining a trade-off between sufficient update signals and knowledge preservation, and ignore computational consumption of the adaptation for large models.}



To this end, we propose a novel TTA framework for foundation medical segmentation models like MedSAM. Compared to existing methods, one of the most pivotal differences is that \textit{our proposed method updates latent representations rather than any model parameters\footnote{We refer to partial or whole model updates as parametric updates hereafter}}, which can not only endow maximal update flexibility under a forgetting-free context but also circumvent heavy forward and backward computations on sophisticated architectures (\textit{e.g.,} image encoder). Specifically, by incorporating an off-the-shelf conditionally independent assumption between the input image $\mathbf{X}$ and the prediction $\mathbf{Y}$ in MedSAM, we provide a theoretical analysis to uncover that directly refining the image embedding $\mathbf{Z}$ (\textit{a.k.a.} latent refinement) can encourage maximizing factorized conditional probabilities  (including latent fidelity $P(\mathbf{Z}|\textbf{X})$  and segmentation likelihood $P(\mathbf{Y}|\mathbf{Z})$) of posterior prediction probability $P(\mathbf{Y}|\mathbf{X})$, in a similar objective like parametric updates. Under this latent refinement framework, we devise a novel distribution-approximated latent conditional random field (CRF) loss to leverage useful cues in the input space for explicitly maximizing the posterior distribution of the latent fidelity.  Meanwhile, we utilize entropy minimization to enforce the usability of the final predictions for implicitly maximizing segmentation likelihood.  Finally, the mask decoder can directly impose the refined latent without inferring the image embedding again. Our contributions include three-fold:
\begin{itemize}
    \item We provide theoretical analyses to uncover that direct latent refinement is feasible to approach the same goal as parametric updates under the MedSAM architecture, which enables us to realize high computational efficiency and segmentation performance (refers to Table \ref{gflops}) without the risk of catastrophic forgetting. 
    \item A  distribution-approximated latent CRF loss based on a Maximum Mean Discrepancy-aware co-occurrent measure is proposed to introduce potential useful cues from input space, leading to sufficient statistics of image embedding over the input image.
    \item The latent refinement framework is validated on three multi-center medical image segmentation datasets with strong appearance variations and simulated prompt shifts. Extensive experiments demonstrate that our method can not only achieve average $\sim$\textbf{3\%} Dice score improvements among three datasets but also enjoy around \textbf{7}$\times$ computational complexity reduction compared with existing methods.
\end{itemize}

\section{Related works}
\label{Methodology}
\subsection{Medical Image Segmentation with TTA}
Various test-time adaptation (TTA) methodologies have been developed to address distribution shifts during inference~\citep{wang2020tent,niu2022efficient,darestani2022test}.  Recent advancements have extended TTA frameworks to more complex tasks, including super-resolution \citep{deng2023efficient} and blind image quality assessment \citep{roy2023test}.
In the field of medical image segmentation, TTA tackles the common multi-center issue~\citep{liu2023fedcl}, where different hospitals may have specific imaging parameters with strong appearance variations~\citep{wang2020dofe}. To this end, some methods conducted the optimization of the BN layer, manipulating either the affine-transformation parameters~\citep{zhang2024pass} or source-trained BN statistics~\citep{dong2024medical}, where many self-supervised objectives and sophisticated alignment of source-trained statistics in the BN layer are leveraged. Meanwhile, some approaches~\citep{wen2024denoising} fine-tune the overall model to explore more gains using elaborated objectives and knowledge accumulation strategies. We argue that such parametric updates suffer from restricted effectiveness due to limited BN-centric update signals and catastrophic forgetting for a large pre-trained model, respectively. The computational complexity is also ignored during adaptation. There are limited works specific to the TTA of foundation medical segmentation models. For example, \citet{huang2024improving} proposed an auxiliary online learning (AuxOL) and adaptive fusion for the foundation medical image segmentation model. However, AuxOL requires clinicians' annotations for model training, leading to potential unfeasibility in many scenarios. Beyond the medical image, \citet{schon2024adapting} leveraged the clicked point prompts 
as cues for adaptation, which is inapplicable for MedSAM due to its bounding box-based prompts. 

Thus, there is a lack of a customized approach for the TTA problem of foundation medical segmentation models when considering both effectiveness and efficiency.

\subsection{Latent Refinement for TTA}
Typically, latent refinement is applied in the neural image compression community~\citep{chen2025testtime,djelouah2019content,lv2023dynamic,shen2023dec,tsubota2023universal} to reduce the bit consumption of latent representation at compression time. 
Since the image compression task has a strong self-supervised signal that corresponds to reconstructing an original raw image by an encoder-decoder architecture~\citep{chen2025testtime,djelouah2019content,lv2023dynamic,shen2023dec}, it is more straightforward to leverage the latent refinement technique. Instead, there is no direct-supervised signal for medical image segmentation during inference, leaving unexplored space. To the best of our knowledge, we are the first to introduce the concept of latent refinement to test-time adaptation of medical image segmentation. More importantly, we provide theoretical analyses to uncover that directly refining the latent representation of the input image is feasible to approach the same goal as parametric update-based counterparts, which is not developed by previous literature.

\section{Methodology}
\label{Methodology}
\subsection{Preliminary and Theoretical Analysis}
This paper mainly discusses the common MedSAM architecture in~\citep{ma2024segment} with the bounding box prompt. During the test-time adaptation, the bounding box prompt is provided and fixed but is omitted in the following derivations for notation simplification.   
\\
\textbf{Definition 1.} \textit{(Architecture of MedSAM) Unlike U-Net and its variants~\citep{ronneberger2015u}, MedSAM has no skip connections between an image encoder $E_{\theta}$ parameterized by $\theta$ and a mask decoder $D_{\varphi}$ parameterized by $\varphi$, which enables multiple interactions by reusing the image embedding.} \\
In light of this, the following proposition can be derived:\\
\textbf{Proposition 1.} \textit{Let $\mathbf{X} \in \mathcal{X}$, $\mathbf{Z} \in \mathcal{Z}$, and $\mathbf{Y} \in \mathcal{Y}$ be an input image, its latent embedding, and its corresponding prediction. For a MedSAM model, $\mathbf{Y}$ is strictly conditionally independent of $\mathbf{X}$ given $\mathbf{Z}$, \textit{i.e.,} $\mathbf{Y} \perp \mathbf{X} \mid \mathbf{Z}$.}\\
\textit{Proof:} Once the (deepest) latent embedding $\mathbf{Z}$ is calculated by $\mathbf{Z}=E_{\theta}(\mathbf{X})$, the upsampling function $\mathbf{Y}=D_{\varphi}(\mathbf{Z})$ is not directly related to the input image $\mathbf{X}$ or its any latent variants, leading to $\mathbf{Y} \perp \mathbf{X} \mid \mathbf{Z}$. $\quad \quad  \quad \quad  \qedsymbol$\\
\textbf{Theorem 1.} \textit{(General objective) Under maximum a posteriori (MAP) estimation, the objective of TTA for image segmentation is computing the optimal prediction $\mathbf{Y}^{*}$ by maximizing the posterior probability conditioned on $\mathbf{X}$:}
    \begin{equation}
    \mathbf{Y}^{*}=D_{\varphi^{*}}(E_{\theta^{*}}(\mathbf{X})), \varphi^{*},\theta^{*}=\arg \max_{\theta,\varphi}[P(\mathbf{Y}|\mathbf{X}; \theta,\varphi)]\footnote{In practice, one would optimize the expectation of the probability. To be simplified, we omit it in later formulations.  }.\nonumber
\end{equation}Various TTA strategies are regarded as implicit posterior maximization by self-supervised loss-based partial or overall parameter updates. 
Here, we incorporate the architecture of the MedSAM as a factorized precondition to yield an extension of Theorem 1 as follows.\\
\textbf{Corollary 1.} \textit{(Factorized objective of MedSAM) The log-posterior $\log P(\mathbf{Y}|\mathbf{X})$ during test-time adaptation can be factorized into the sum of the segmentation likelihood $\log P(\mathbf{Y}|\mathbf{Z})$ and
the latent fidelity term $\log P(\mathbf{Z}|\mathbf{X})$ based on $\mathbf{Y} \perp \mathbf{X} \mid \mathbf{Z}$, \textit{i.e.,}}
    \begin{gather}
    \mathbf{Y}^{*} = D_{\varphi^{*}}(\mathbf{Z}^{*}), \mathbf{Z}^{*}=E_{\theta^{*}}(\mathbf{X})  \nonumber\\  
    \varphi^{*}, \theta^{*} = \arg \max_{\varphi,\theta}[\log P(\mathbf{Y}|\mathbf{Z};\varphi) + \log P(\mathbf{Z}|\mathbf{X};\theta)],\nonumber
\end{gather}
\textit{Proof:} By recalling Prop. 1, $P(\mathbf{Y}|\mathbf{X}) \propto P(\mathbf{Y}|\mathbf{Z})\cdot P(\mathbf{Z}|\mathbf{X})$ holds. Substituting this into the MAP objective in Theorem 1 yields the factorized objective above. Please refer to the Appendix for more details. $\quad  \quad \quad  \qedsymbol$ 
\begin{figure*}[!h]
    \centering
    \includegraphics[width=0.95\linewidth]{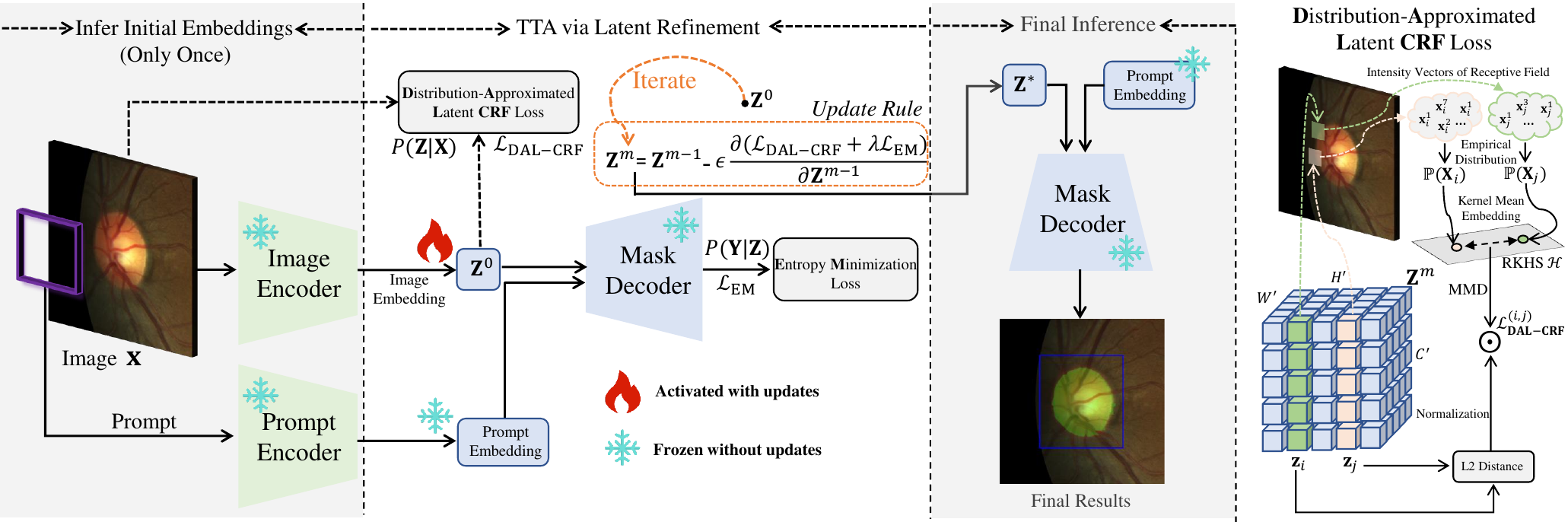}
    \caption{The framework of the proposed method, including initial embedding inference, TTA via latent refinement, and final inference phases. Distribution-approximated latent CRF loss is visualized, where the loss between $\mathbf{z}_{i}$ and $\mathbf{z}_{j}$ is computed. }
    \label{framework}
\end{figure*}
\vspace{-0.35cm}
\subsection{Latent Refinement in MedSAM framework}
\textbf{Problem.} For Corollary 1, one of the commonly adopted solutions for achieving optimal prediction $\mathbf{Y}^{*}$ is directly to jointly optimize partial or whole parameters in \{$\varphi$,$\theta$\}, but two challenges are encountered as follows.
\begin{itemize}
    \item \textit{Partial model updates.} Most existing approaches are BN-centric strategies, which are inapplicable for LN-based MedSAM. Although updating the affine-transformation parameters in LN layers is feasible, the performance may be marginal due to restricted model modifications especially in strong domain shifts.
    \item \textit{Whole model updates.} It is difficult to maintain a trade-off between TTA performance and catastrophic forgetting. Meanwhile, optimizing the overall model is computationally expensive due to the considerable number of parameters in modern large models. 
\end{itemize}
\textbf{Solution.} In light of this, we are interested in directly optimizing the latent embedding rather than the model parameters. Assume the encoder $E_{\theta}$ and the decoder $D_{\varphi}$ are fixed during TTA, this scheme can not only circumvent the above-mentioned challenges (\textit{i.e.,}  \textbf{it is likely to update more ``parameters'' in the forgetting-free context}) but also be equivalent to theoretically achieving the same factorized objective of MedSAM as Corollary 1,
\begin{gather}
     \mathbf{Y}^{*} = D_{\varphi}(\mathbf{Z}^{*}),  \nonumber\\  
    \mathbf{Z}^{*} = \arg \max_{\mathbf{Z}}[\log P(\mathbf{Y}|\mathbf{Z};\varphi) + \log P(\mathbf{Z}|\mathbf{X};\theta)], \label{optimization objective}
\end{gather}where $\mathbf{Z}=E_{\theta}(\mathbf{X})$ can be initialized and refined iteratively. Typically, the total parameter number of $\mathbf{Z}$ is overly larger than that of the normalization layer, resulting in more update flexibility. Under frozen $\theta$ and $\varphi$, there is no knowledge forgetting during optimization. These two perspectives are more likely to activate better TTA performance collaboratively than existing methods. \\
\textbf{Effectiveness of Latent Refinement during TTA.} First, the $\log P(\mathbf{Z}|\mathbf{X})$ term encourages $\mathbf{Z}$ to maintain sufficient statistics over the input $\mathbf{X}$, which is highly significant for the MedSAM framework during the TTA because 1) domain-specific variations at test time may degrade salient characteristics (\textit{e.g.,} segmentation cues like obvious edges or contours) of latent embedding, leading to suboptimal segmentation performance; 2) sufficient statistics relative to the input $\mathbf{X}$ can encourage the conditional independence assumption of MedSAM,  \textit{i.e.,} $\mathbf{Y} \perp \mathbf{X} \mid \mathbf{Z}$, to hold. Second, $\log P(\mathbf{Y}|\mathbf{Z})$ guarantees the usability of latent embedding for final predictions, preventing any collapsed solutions.\\
\textbf{Practical implementation.} For Eq. (\ref{optimization objective}), one would update the initial latent embedding $\mathbf{Z} = \mathbf{Z}^{0}=E_{\theta}(\mathbf{X})$ with $m\ge1$ steps,  for each step $\mathbf{Z}^{m} = $
    \begin{equation}
    \mathbf{Z}^{m-1} - \epsilon\cdot \frac{\partial[-\log P(\mathbf{Y}^{m-1}|\mathbf{Z}^{m-1};\varphi) - \log P(\mathbf{Z}^{m-1}|\mathbf{X};\theta]}{\partial \mathbf{Z}^{m-1}}\label{optimization details}
\end{equation}where $\epsilon$ is the learning rate. The final prediction can be represented as $\mathbf{Y}^{m} = D_{\varphi}(\mathbf{Z}^{m})$. 
\subsection{Distribution-Approximated Latent CRF}
To Eq. (\ref{optimization details}), one of the most important components is maximizing the log-probability $P(\mathbf{Z}^{m-1}|\mathbf{X};\theta)$, \textit{i.e.,} the latent fidelity term. For MedSAM, the volume of the image encoder parameter $\theta$ is considerable, resulting in a high computation during the forward and backward propagations in the TTA phase. Benefiting from our latent refinement framework, we propose circumventing the image encoder $\theta$ to directly enforce sufficient statistics over the input image $\mathbf{X}$, \textit{i.e.,} maximizing $\log P(\mathbf{Z}^{m-1}|\mathbf{X})$. By doing so, we can leverage potential useful cues in the input space, which is typically ignored by existing TTA methods that focus more on the label space (\textit{e.g.,} prediction consistency~\citep{chen2024gradient}).

Therefore, we introduce the concept of conditional random field (CRF)~\citep{sutton2012introduction}, which is usually employed to refine the final predictions by enforcing the co-occurrence between the label space $\mathcal{Y}$ and the input space $\mathcal{X}$. For our purpose, we extend the CRF to the co-occurrence between the latent space $\mathcal{Z}$ and the input space $\mathcal{X}$. Theoretically, minimizing the CRF energy is equivalent to maximizing the posterior probability of the random field~\citep{krahenbuhl2011efficient}, which coincides with our optimization objective in Eq. (\ref{optimization details}) via a computationally efficient manner. Formally, given a random field $\mathbf{Z}=E_{\theta}(\mathbf{X})$ over a set of vectors \{$\mathbf{z}_{1},\mathbf{z}_{2},\cdots,\mathbf{z}_{N'}$\}, where $\mathbf{Z} \in \mathbb{R}^{W'\times H' \times C'}$ is the latent embedding with a spatial resolution of $N'$=$W'$$\times$$H'$ and a channel of $C'$, \textit{i.e.,} $\mathbf{z} \in \mathbb{R}^{1\times C'}$. Consider another random field $\mathbf{X}$ over a set of vectors \{$\mathbf{x}_{1},\mathbf{x}_{2},\cdots,\mathbf{x}_{N}$\}, where $\mathbf{X} \in \mathbb{R}^{W\times H \times C}$ is the input intensities with a spatial resolution of $N$=$W$$\times$$H$ and a channel of $C$, \textit{i.e.},  $\mathbf{x} \in \mathbb{R}^{1\times C}$. 

Typically, $\mathbf{x}_{i}$ and  $\mathbf{z}_{i}$ are related in a one-to-one manner, \textit{i.e.,} $\mathbf{x}_{i}$ represents a color vector at pixel $i$ in the input space and $\mathbf{z}_{i}$ is its corresponding predicted label in the label space. However, the input and latent spaces are dimensionally heterogeneous, \textit{i.e.,} $N\neq N'$, which makes constructing a random field in this scenario challenging.
Here, a novel distribution-approximated latent CRF (DAL-CRF) is proposed to eliminate the dimensionally heterogeneous problem between the input and latent spaces. Specifically, by leveraging the fact that each latent vector $\mathbf{z}_{i}$ corresponds to a receptive field in the input space with a set of intensity vectors, it is feasible to identify related intensity vectors for each latent vector, \textit{i.e.,}
    \begin{equation}
   \textbf{z}_{i} \rightarrow \mathbf{X}_{i} = \{\mathbf{x}_{i}^{l}\}_{l=1}^{L}, i = 1,2,\cdots, N',
\end{equation}where $L$ denotes the number of intensity vectors in the receptive field, where $L=256$ corresponding to a $16\times 16$ receptive field in MedSAM. Then, by following \citet{veksler2023test}, the pairwise potential is only considered to be a regularization loss to encourage the minimization of CRF energy, which can be formulated as $\mathcal{L}_{\operatorname{DAL-CRF}} = $
    \begin{equation}
    \frac{1}{N'}\sum_{i,j \in \mathcal{G}} (\frac{1}{2}\operatorname{MMD}(\mathbb{P}(\mathbf{X}_{i}),\mathbb{P}(\mathbf{X}_{j})))\cdot \exp(-\|\mathbf{z}_{i}-\mathbf{z}_{j}\|_{F}). \label{CRF}
\end{equation}Unlike existing CRF methods~\citep{veksler2023test}, which utilize the Gaussian kernel to evaluate the correlation between two color vectors,  we introduce the Maximum Mean Discrepancy (MMD) to measure the distance between two vector sets. Specifically, each vector set is first approximated as an \textit{empirical distribution}, \textit{i.e.,} $\mathbf{x}_{i}^{l} \sim \mathbb{P}(\mathbf{X}_{i})$. Then, each empirical distribution is represented by a \textit{kernel mean embedding} in a reproducing kernel Hilbert space (RKHS) $\mathcal{H}$, \textit{i.e.,} $\mu_{\mathbb{P}}=\mathbb{E}_{\mathbf{x}\sim\mathbb{P}}[\phi(\mathbf{x})]=\int_{\mathcal{X}}k(\mathbf{x},\cdot)d\mathbb{P}=\mathbb{E}_{\mathbf{x}\sim\mathbb{P}}[k(\mathbf{x},\cdot)]$, where $k:\mathcal{X}\times\mathcal{X}\rightarrow \mathbb{R}$ is a reproducing kernel (Gaussian RBF kernel here) and corresponding feature map $\phi:\mathcal{X}\rightarrow \mathcal{H}$.  The discrepancy can be measured between mean embeddings, \textit{i.e.,}
    \begin{equation}
    \operatorname{MMD}(\mathbb{P}(\mathbf{X}_{i}),\mathbb{P}(\mathbf{X}_{j})) = \| \frac{1}{L}\sum_{l=1}^{L}\phi(\mathbf{x}_{i}^{l})-\frac{1}{L}\sum_{l=1}^{L}\phi(\mathbf{x}_{j}^{l})\|_{\mathcal{H}}^{2}.
\end{equation} MMD enables all-order correlation measurement using Gaussian kernel via unbiased empirical estimation. This may be useful for better discrepancy measurement, as the distribution of the intensity space may be highly complex and non-linear.
For the second term in Eq. (\ref{CRF}), we utilize the Frobenius norm $\|\cdot\|_{F}$ to calculate the feature compatibility in a co-occurrent probability using exponential zero-to-one normalization.\\
\textbf{Integrating Bounding Box Prompt.} The graph $\mathcal{G}=\{\mathcal{V},\mathcal{E}\}$ in Eq. (\ref{CRF}) is usually either a complete (dense-connected) graph~\citep{krahenbuhl2011efficient} or a sparse position-related neighboring graph~\citep{veksler2023test} on $\mathbf{Z}$. Instead, we propose to consider the bounding box prompt as the prior to construct $K$-random pairs inside the given bounding box, which can not only exclude those unimportant latent features far away from the bounding box prompt to enhance the computational efficiency but also focus more on the interested region with better performance.\\
\textbf{Intuitive Understanding.} For Eq. (\ref{CRF}), one actually encourages being more discriminative in latent representations between $\mathbf{z}_{i}$ and $\mathbf{z}_{j}$. Since if two empirical distributions have a large distribution discrepancy (\textit{i.e.,} the MMD distance is large) in the input space, there are more penalties assigned to the pair who has larger similarity (\textit{i.e.,} co-occurrent probability is larger)) in the latent space. By minimizing $\mathcal{L}_{\operatorname{DAL-CRF}}$, one can enforce the latent representations to equip with sufficient statistics as in the input space.  
\subsection{Overall Objective at Test-time Adaptation}
Without accessing the ground-truth label, it is infeasible to maximize the segmentation likelihood $\log P(\mathbf{Y}^{m}|\mathbf{Z}^{m};\varphi)$ directly. Therefore, we adopt a common predictive entropy minimization to implicitly achieve this goal~\citep{grandvalet2004semi}. 
Formally,
\begin{gather}
\mathcal{L}_{\operatorname{EM}} = H(P(\mathbf{Y}|\mathbf{Z})) = -\sum_{h,v\in C}P(y^{h,v}|\mathbf{Z})\log P(y^{h,v}|\mathbf{Z}),\nonumber\\
    \operatorname{where} \quad C=\{(h,v)|P(y^{h,v}|\textbf{Z})\ge \alpha\}, \label{entropy}
\end{gather}
where $y^{h,v}$ denotes a single prediction in the position $(h,v)$ of $\mathbf{Y}$. Eq. (\ref{entropy}) is similar to foreground-background balanced entropy loss in \citet{dong2024medical}. However, we only utilize the foreground entropy and set a high confidence threshold (\textit{e.g.,} $\alpha=0.95$) due to the following reasons. First, since the CRF-based loss usually requires a volumetric regularization to avoid the trivial solutions~\citep{veksler2023test} (\textit{e.g.,} all labels are assigned to background), the entropy regularization over a high confidence foreground region can be interpreted as an implicit volumetric regularization, \textit{i.e.,} ensuring these foreground pixels confident enough for preventing trivial solutions. Second, we argue that background entropy is unnecessary for a foundation segmentation model, as these models have learned a good foreground-background balance.

\begin{table*}[!ht]
    \renewcommand\arraystretch{0.95}
    \centering
   
    \resizebox{0.99\textwidth}{!}{%
    \setlength\tabcolsep{6.0pt}
    \scalebox{1.00}{
    \begin{tabular}{c|c|c c  c  c  c||c c  c c c}
        \hline \hline
        \multirow{2}{*}{Update Type} & \multirow{2}{*}{Method} & \multicolumn{5}{c||}{\textbf{Dice Coefficient~$\uparrow$}} & \multicolumn{5}{c}{\textbf{Average Surface Distance~$\downarrow$}}\\ 
        \cline{3-12}
        & & D1 & D2 & D3 & D4 & Avg. & D1 & D2 & D3 & D4 & Avg.\\ 
        \hline \hline
      \rowcolor{blue!10} Looseness prompts   &    Direct Inference & 90.99	&87.16&	87.06&	84.72	&87.48&	9.48	& 7.51	& 8.74	& 7.62	& 8.33

 \\ 
        \hline
        \multirow{4}{*}{Normalization update} 
         & TENT~(ICLR'21) & \underline{91.27}	&85.40	& 85.06	& 81.84&	85.89&	9.09&	8.70& 	10.05&	8.88&	9.18

  \\ 
         & InTEnt~(CVPR'24)  &  \textbf{91.36}	& \underline{87.43}	& \underline{87.47}	& \underline{86.13}&	87.60&
	\textbf{8.80} & \underline{7.51}	& \underline{8.26}	&7.66 	& 8.06

  \\ 
         & GraTa~(AAAI'25)   &  90.89	&87.00&	86.95&	84.64	& 87.37
&	9.48&	7.54&	8.80&	\underline{7.65} &	8.37

  \\ 
         & PASS~(TMI'24)     &  80.86	& 79.63	& 82.03	& 84.02&	81.63&	20.40&	12.22&	11.69&	7.97&	13.07

  \\ 
        \hline
        \multirow{2}{*}{Full updates} 
         & MEMO~(NeurIPS'22)  &  87.26	&85.48	& 86.42	& 84.47&	85.91&	12.45&	8.27&	9.15&	7.76&	9.40
  \\ 
         & CRF-SOD~(CVPR'23)  &  89.57 &	84.01	& 84.80&	82.58& 	85.24&	10.54&	8.99&	10.18&	8.64&	9.58
  \\ 
        \hline
         \rowcolor{gray!10} Latent updates &   \textit{Ours} &89.35	& \textbf{88.73}	& \textbf{89.83}	& \textbf{88.18}	& \textbf{89.02}	& 11.40	& \textbf{6.69}& 	\textbf{6.92} &	\textbf{6.07}&	\textbf{7.77}

  \\ 
        \hline \hline
        \rowcolor{blue!10} Shrinkage prompts   &  Direct Inference & 81.05	&67.60	& 72.10	& 68.57&	72.33	& 18.51	& 17.61	& 17.76	& 14.92	& 17.20

 \\ 
        \hline
        \multirow{4}{*}{Normalization update} 
         & TENT~(ICLR'21) &  80.22	& 65.17	& 69.25	&62.10&	69.18 & 18.85	& 19.26& 	19.39	& 17.58	& 18.77

  \\ 
         & InTEnt~(CVPR'24)  & 82.72	 & 70.06	& 74.50	&73.49& 	75.19&	16.86&	15.28&	16.13&	13.87&	15.53

  \\ 
         & GraTa~(AAAI'25)   &  80.87	& 67.32	& 71.94&	68.46	& 72.15	& 17.27&	16.54&	16.53&	14.52&	16.21

  \\ 
         & PASS~(TMI'24)     &  66.81	& 58.66	& 63.31	& 56.78	& 61.39	& 32.39&	22.89	& 22.74	& 20.19	& 24.55

  \\ 
        \hline
        \multirow{2}{*}{Full updates} 
         & MEMO~(NeurIPS'22)  &  81.02	& 67.54	& 72.10&	68.84&	72.37&	18.52&	17.76&	17.77&	14.78&	17.20

  \\ 
         & CRF-SOD~(CVPR'23)  &  81.84	& 69.26	& 74.73	&72.08&	74.48&	19.27	& 18.83&	18.63&	14.73&	17.86

  \\ 
        \hline
        \rowcolor{gray!10} Latent updates &   \textit{Ours}  & \textbf{83.07} &\textbf{73.72}& \textbf{78.54}& \textbf{75.77}& \textbf{77.77}& \textbf{16.74}& \textbf{14.38}& \textbf{13.72}& \textbf{11.26}& \textbf{14.02}

  \\ 
        \hline \hline
        \rowcolor{blue!10}  Perfectness prompts   & Direct Inference & 87.58	& 80.25	& 81.32	& 78.46	&81.90

& 12.76& 	11.21	& 12.27	& 10.24	& 11.62

 \\ 
        \hline
        \multirow{4}{*}{Normalization update} 
         & TENT~(ICLR'21) & \textbf{87.60}	&77.98	& 78.18&	72.99&	79.18
	& \underline{12.83}	& 12.21	& 13.92	& 12.69	& 12.91

  \\ 
         & InTEnt~(CVPR'24)  &  86.54 &	84.25	& 85.17	& 81.65	& 84.40
	& \textbf{10.98}	& 9.23&	9.85&	8.95&	9.75

  \\ 
         & GraTa~(AAAI'25)  &  \underline{87.39}	& 79.97	& 81.14	& 78.36	& 81.71& 
	12.96	& 11.38	& 12.38	& 10.29	& 11.75
\\ 
         & PASS~(TMI'24)     &  74.59 &	69.86&	73.88	& 71.34	& 72.41
& 25.53	& 17.32	& 16.58	& 13.85	& 18.32

 \\ 
        \hline
        \multirow{2}{*}{Full updates} 
         & MEMO~(NeurIPS'22)  & \underline{87.39}	& 80.18	& 81.32	& 78.49 &	81.84&
	12.97	& 11.24	&12.27& 	10.23	& 11.67

  \\ 
         & CRF-SOD~(CVPR'23)  & 86.07 & 	77.22	& 79.10	& 76.24	&79.65
	& 13.96	& 12.59	& 13.63	& 11.2	& 12.84

  \\ 
        \hline
       \rowcolor{gray!20} Latent updates &   \textit{Ours}  & 86.85	 & \textbf{85.11}	& \textbf{86.72}	& \textbf{84.49}	& \textbf{85.79}
& 13.73	& \textbf{8.63}	& \textbf{8.79}	& \textbf{7.48}	& \textbf{9.65}

 \\ 
        \hline \hline
    \end{tabular}
    }}
    \caption{Quantitative comparison of TTA results on fundus image \textbf{OC segmentation} between different methods. The best and second-best results are shown in the \textbf{bold} and the \underline{underline}, respectively. For loose prompts, we report the average of two looseness ratios. For shrinkage prompts, we report the average of two shrinkage ratios. Full results can be found in the Appendix.}    
\label{tab:comparisons_fundus_OC}
\vspace{-0.4cm}
\end{table*}
By considering the proposed distribution-approximated latent CRF loss and the customized entropy loss, the practical latent refinement scheme is reformulated as follows.
\begin{equation}
    \mathbf{Z}^{m} = \mathbf{Z}^{m-1} - \epsilon\cdot \frac{\partial[\mathcal{L}_{\operatorname{DAL-CRF}}+\lambda\mathcal{L}_{\operatorname{EM}}]}{\partial \mathbf{Z}^{m-1}},\label{optimization details 2}
\end{equation}
where $\lambda$ is a balance coefficient. 
\subsection{Implementation Details}
The overall framework is illustrated in Figure \ref{framework}. Superficially, the initial image embedding $\mathbf{Z}^{0}$ is inferred only once. Once the refined latent representation $\mathbf{Z}^{*}$ is obtained, the upsampling can be conducted directly without requiring image embedding again. This mechanism is well compatible with the interactive scheme of the MedSAM. We implement the proposed method based on official MedSAM checkpoints (ViT-B pre-trained models) and codes using a one-step optimization strategy ($m$=1) by following previous TTA settings. Due to the extremely various sizes of RoIs in medical images, we leverage the initial segmentation result $\mathbf{Y}^{0}$ and the given bounding box prompt to explore a dynamic adjustment strategy for important hyperparameters. Specifically, the learning rate $\epsilon$ with the Adam optimizer is linearly interpolated from $5\times 10^{2}$ to $1\times10^{-1}$ according to the ratio of the initial foreground to the entire image $f_{e}$ (\textit{i,e.,} the rough size of the RoI), where $f_{e}\le0.05$ and $f_{e}\ge0.2$ are fixedly set to $5\times 10^{2}$ and $1\times10^{-1}$, respectively. Note that we use commonly-used definitions over the $f_{e}$ as adopted by \citet{GAO2021101831}, \textit{i.e.,} $f_{e}\le0.05$, $0.05<f_{e}<0.2$, and $f_{e}\ge0.2$ as small, medium, and large RoIs. Similarly, the $\lambda$ is also RoI-specific adjustment, where the small RoI ($f_{e}\le0.05$) requires a similar contribution between the $\mathcal{L}_{\operatorname{DAL-CRF}}$ and the $\mathcal{L}_{\operatorname{EM}}$, thus the $\lambda$ is set to $1\times10^{3}$. 
The $K$  is set to 4 by balancing computational consumption and performance (refers to Table \hyperref[tab:TRAIN parameter]{5(a)}). $\alpha$ in entropy loss is set to 0.95 emprically. 
\begin{table*}[!ht]
    \renewcommand\arraystretch{0.95}
    \centering
   
    \resizebox{0.99\textwidth}{!}{%
    \setlength\tabcolsep{6.0pt}
    \scalebox{1.00}{
    \begin{tabular}{c|c|c c  c  c  c||c c  c c c}
        \hline \hline
        \multirow{2}{*}{Update Type} & \multirow{2}{*}{Method} & \multicolumn{5}{c||}{\textbf{Dice Coefficient~$\uparrow$}} & \multicolumn{5}{c}{\textbf{Average Surface Distance~$\downarrow$}}\\ 
        \cline{3-12}
        & & D1 & D2 & D3 & D4 & Avg. & D1 & D2 & D3 & D4 & Avg.\\ 
        \hline \hline
      \rowcolor{blue!10}  Looseness prompts   &  Direct Inference & 83.05	&78.86	& 81.82	&75.54	&79.81	& 33.20&	32.87&	30.29&	31.90&	32.06
 \\ 
        \hline
        \multirow{4}{*}{Normalization update} 
         & TENT~(ICLR'21) &  81.61 &	76.96	& 80.00	& 74.85	& 78.35	&37.58	& 36.54&	33.53	& 34.35	& 35.50
  \\ 
         & InTEnt~(CVPR'24)  &  83.61	& 76.71	&80.00	& 71.83	& 78.03	& 32.46&	37.01	& 32.57	& 39.14	& 35.29
  \\ 
         & GraTa~(AAAI'25)   &  84.64	& 80.04	& 84.10	& \textbf{82.82} &	82.90	& 30.02&	30.93	& 25.91	&28.91	& 28.94
  \\ 
         & PASS~(TMI'24)     &  84.57	&81.77&	82.29&	77.34	& 81.49
&	18.02	&26.20&	21.09	&24.68	&22.49
  \\ 
        \hline
        \multirow{2}{*}{Full model updates} 
         & MEMO~(NeurIPS'22)  &  86.26&	80.01&	83.94	&78.18	&82.10&	26.73&	30.93&	26.27	&28.54&	28.12  \\ 
         & CRF-SOD~(CVPR'23)  &  87.03 &	82.86&	83.33&	79.52	& 83.18&	25.19&	26.25	& 23.72&	26.53	&25.42   \\
        \hline
        \rowcolor{gray!10} Latent updates &   \textit{Ours}  &  \textbf{90.85}	& \textbf{83.55}&	\textbf{87.77}	& \underline{82.24}	& \textbf{86.10} &	\textbf{17.59} &	\textbf{ 23.54}	& \textbf{18.50} &	\textbf{21.75} &	\textbf{20.34}
  \\ 
        \hline \hline
         \rowcolor{blue!10} Shrinkage prompts   & Direct Inference & 92.70 &	86.17	& 90.90	& 88.67	& 89.61&	12.60&	17.66	& 12.77	& 12.10	& 13.78
 \\ 
        \hline
        \multirow{4}{*}{Normalization update} 
         & TENT~(ICLR'21) &  90.63& 	85.63	& 89.66	& 86.88	& 88.20 &	15.90	& 18.44	& 14.27	& 12.48	& 15.27
  \\ 
         & InTEnt~(CVPR'24)  & 90.91	& \textbf{87.27}	& 90.88& 	88.70&	89.44	& 15.49	& 16.47&	12.68&	12.05	&14.17
  \\ 
         & GraTa~(AAAI'25)   &  92.61	& 85.60	& 91.00	&88.71&	89.48&	12.70&	18.32&	12.61&	12.11&	13.93
  \\ 
         & PASS~(TMI'24)     &  91.46 &	78.85	& 85.92	& 75.70	& 82.98	& 13.72	& 25.14&	17.54&	23.05&	19.86
  \\ 
        \hline
        \multirow{2}{*}{Full model updates} 
         & MEMO~(NeurIPS'22)  &  92.12	& 83.82	&90.87	&88.77	&88.89	&13.52	&20.18	& 12.75	& 12.04	& 14.62
  \\ 
         & CRF-SOD~(CVPR'23)  &  \textbf{94.42}	& 84.07 &	92.34	& 91.43	& 90.56	& 9.36	&19.83	& 10.70	& 9.30	& 12.30
  \\ \hline
       \rowcolor{gray!10} Latent updates &   \textit{Ours}  &  \underline{94.24} &	\underline{86.77}&	\textbf{93.23}	& \textbf{92.42} &	\textbf{91.66} &	\textbf{9.80}	& \textbf{16.25}	& \textbf{9.45}	& \textbf{8.21}	& \textbf{10.93}
  \\ 
        \hline \hline
      \rowcolor{blue!10}  Perfectness prompts   &  Direct Inference & 96.49 & 92.87
  &  93.89
  &  92.36
 & 93.90
 &  6.09
 &  9.48
 &  8.64
 & 8.36
  & 8.14
 \\ 
        \hline
        \multirow{4}{*}{Normalization update} 
         & TENT~(ICLR'21) &  94.20 &	90.78	&93.39	&90.38&	92.18	&10.12 &	12.01	&9.32	&10.39	&10.46
  \\ 
         & InTEnt~(CVPR'24)  &  94.12 &	91.57 &	94.42	& 93.47	& 93.39	&10.12&	11.01&	8.01	&7.03	&9.04
  \\ 
         & GraTa~(AAAI'25)  &  94.46	&92.66&	93.91&	92.31&	93.33&	6.14	&9.73&	8.61	&8.42	& 8.22\\ 
         & PASS~(TMI'24)     &  94.90 &	85.35&	84.36	&86.14	&87.68&	8.58	&17.52&	17.62	&13.98&	14.42
 \\ 
        \hline
        \multirow{2}{*}{Full model updates} 
         & MEMO~(NeurIPS'22)  & 95.77 &	91.48	&93.89&	92.31&	93.36&	7.80	& 11.28	&8.61	&8.41	&9.02
  \\ 
         & CRF-SOD~(CVPR'23)  &  96.66	& 91.15&	93.76	&93.41&	93.75	&5.78&	11.42&	8.73&	7.26&	8.29
  \\
        \hline
        
         \rowcolor{gray!10} Latent updates &   \textit{Ours}  &\textbf{96.68} & \textbf{92.67}
 &  \textbf{95.15}
&\textbf{94.57}
  &\textbf{94.76}
  & \textbf{5.69}
 & \textbf{9.51}
 & \textbf{6.85}
 &  \textbf{5.98}
& \textbf{7.00}
 \\
        \hline \hline
    \end{tabular}
    }}
     \caption{Quantitative comparison of TTA results on fundus image \textbf{OD segmentation} between different methods. The best and second-best results are shown in the \textbf{bold} and the \underline{underline}, respectively. For loose prompts, we report the average of two looseness ratios. For shrinkage prompts, we report the average of two shrinkage ratios. Full results can be found in the Appendix.}
\label{tab:comparisons_fundus_OD}
\vspace{-0.3cm}
\end{table*}
\begin{table*}[!ht]
    \renewcommand\arraystretch{0.95}
    \centering
   
    \resizebox{0.99\textwidth}{!}{%
    \setlength\tabcolsep{6.0pt}
    \scalebox{1.00}{
    \begin{tabular}{c|c|c c  c  c  c||c c  c c c}
        \hline \hline
        \multirow{2}{*}{Update Type} & \multirow{2}{*}{Method} & \multicolumn{5}{c||}{\textbf{Dice Coefficient~$\uparrow$}} & \multicolumn{5}{c}{\textbf{Average Surface Distance~$\downarrow$}}\\ 
        \cline{3-12}
        & & D1 & D2 & D3 & D4 & Avg. & D1 & D2 & D3 & D4 & Avg.\\ 
        \hline \hline
       \rowcolor{blue!10}      &  Direct Inference & 55.35 &	54.20	&61.54	& 56.67	& 56.94	& 10.22& 	11.39	& 15.27	& 15.95	& 13.20

 \\ 
        \hline
        \multirow{4}{*}{Normalization update} 
         & TENT~(ICLR'21) & 49.66 &	52.96	& 59.39	& 55.38	& 54.34 &	9.94	& \textbf{10.43}	& 15.85	& 16.12	& 13.08

  \\ 
         & InTEnt~(CVPR'24)  &  50.91	& 52.12	& 45.85	& 50.38	& 49.81&	9.35	&10.99	&22.57	& 19.01	& 15.48

  \\ 
         & GraTa~(AAAI'25)  &  56.35	& 55.14	& 63.66	& 58.12	& 58.31&	9.26	& \underline{10.44}	& 14.36	& 15.15	& 12.30

\\ 
         & PASS~(TMI'24)     &  55.24 &	54.80	& 61.94	& 52.88	& 56.21 &	10.58	& 11.01	& 15.02	& 20.14	& 13.99

 \\ 
        \hline
        \multirow{2}{*}{Full updates} 
         & MEMO~(NeurIPS'22)  &56.20 & 	55.20 &	61.20&	57.34&	57.48&	9.27&	10.45&	14.89&	15.39&	12.50

  \\ 
         & CRF-SOD~(CVPR'23)  &54.20	& 54.08	& 61.89	& 57.07& 	56.81&	10.37&	10.83&	15.36&	15.58	& 13.03

  \\ 
        \hline
       \rowcolor{gray!10} Latent updates &   \textit{Ours}  & \textbf{57.92} & \textbf{55.42}& \textbf{66.00} & \textbf{58.62} & \textbf{59.49} & \textbf{8.35} & 10.58 & \textbf{14.14}& \textbf{14.15}&   \textbf{11.75}

 \\ 
        \hline \hline
    \end{tabular}
    }}
    \caption{Quantitative comparison of TTA results on MRI image \textbf{SCGM segmentation} between different methods. SCGM segmentation only reported PP-based results due to its very challenging intricate structures and appearance perturbations. }
\label{tab:comparisons_spin}
\vspace{-0.3cm}
\end{table*}
\section{Experiments}
\textbf{Datasets and Evaluation Metrics}
We evaluate our method on two public multi-center medical image segmentation datasets with different data modalities. (1) \textbf{Fundus} \cite{wang2020dofe}: Fundus datasets are retinal fundus images collected from 4 healthcare centers (D1 to D4 for short). For each retinal fundus image, there are two segmentation objectives, including the optic cup (OC) (mainly small RoIs) and optic disc (OD) (mainly medium or large RoIs). (2) \textbf{Spinal Cord Gray Matter (SCGM)} \cite{li2020domain}: SCGM datasets are collected from 4 institutions (D1 to D4 for short) with different imaging manufacturers for spinal MRI image gray matter segmentation. We use the Dice Similarity Coefficient (DSC)
and Average Surface Distance (ASD) to measure the accuracy.

\begin{table*}[!t]

	\begin{minipage}[t][][c]{0.64\linewidth}
 \vspace{0pt}
		\centering
		\includegraphics[width=\textwidth,height=5.3cm]{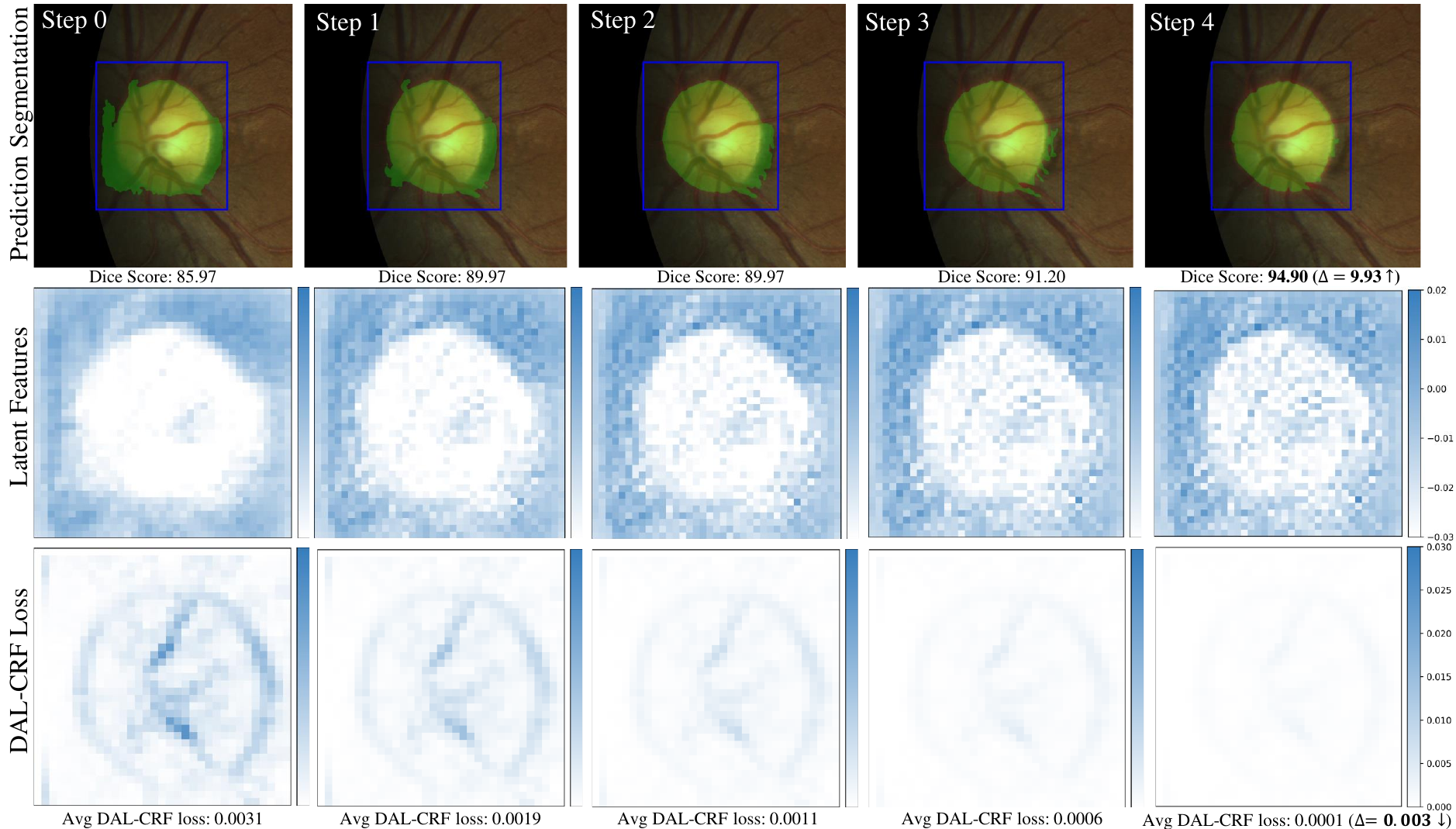}
            \vspace{-0.7cm}
		\captionof{figure}{Visulized process of latent refinement with iterations. The first, second, and third rows denote the segmentation result, corresponding visualization of channel-average refined latent embedding (which is from the region in bounding box prompt and has been fused with the prompt for a better view), and corresponding DAL-CRF loss for each spatial position. You may zoom in for a better view. } 
		\label{visualization latent}
	\end{minipage}  \thinspace \hfill 
	\begin{minipage}[t][][c]{0.36\linewidth}
 \vspace{-0.1cm}
\centering
	
	\label{tab:TRAIN parameter}
        \vspace{-0.15cm}
        \renewcommand{\arraystretch}{1.1}
		\begin{adjustbox}{width=0.95\textwidth}
            
		\begin{tabular}{c|ccccc}
 \multicolumn{6}{c}{(a)}                                                                                                                  \\ \toprule
$K$ & D1 & D2 & D3& D4 & Avg. \\ \hline
2  & 93.24 & 87.06 & \textbf{92.51} & 86.51 & 89.75 \\
4  & 94.16 & \textbf{87.16} & 91.83 & 87.57 & 90.14 \\
6  &   94.67	& 87.08&92.34 &	87.92	& \textbf{90.37}
\\
8  & \textbf{94.86}	& 86.54	&90.81 &	\textbf{88.01}	& 90.03\\
\toprule \toprule
\multicolumn{6}{c}{(b)}  \\ \toprule
  Loss & D1 & D2 & D3 & D4 & Avg. \\ \toprule
DI &  89.19	& 84.31	& 87.92	& 82.39 & 85.95\\ 
$\mathcal{L}_{\operatorname{InTEnt}}$ & 79.34	& 84.89	& 88.83	& 86.08	& 84.78
\\
$\mathcal{L}_{\operatorname{TENT}}$ & 85.09	 & 84.99 &	89.46	& 85.59	& 86.28
 \\
$\mathcal{L}_{\operatorname{CRF-SOD}}$ & 88.93	& 86.17	& 88.42	& 87.33	& 87.71 \\
\rowcolor{blue!10}  $\mathcal{L}_{\operatorname{DAL-CRF}}$ & \textbf{94.16} & \textbf{86.96} &  \textbf{90.85} & \textbf{87.57} & \textbf{89.88}
\\ \toprule
\toprule
\end{tabular}
        \end{adjustbox}
        \caption{ (a) Hyperparameter Analyses in terms of $K$. (b) Integration of the latent refinement framework and other TTA loss functions  \textit{v.s.} Our proposed DAL-CRF, where DI denotes direct inference without adaptation on OD segmentation.}
	\end{minipage}
	\vspace{-0.4cm}
\end{table*}
\begin{table}[t]
\centering  

\resizebox{\columnwidth}{!}{  
\begin{tabular}{l|ccccc}
\hline
Method & D1 & D2 & D3& D4 & Avg. \\
\hline
Direct Inference &  89.19	& 84.31	& 87.92	& 82.39 & 85.95
  \\ 
+ $\mathcal{L}_{\operatorname{EM}}$ & 85.82 & 85.31 & 88.10 & 83.21 & 86.01 \\
+ $\mathcal{L}_{\operatorname{DAL-CRF}}$ & \textbf{94.16} & 86.96 &  90.85& \textbf{87.57} & 89.88 \\
\rowcolor{blue!10} $+\mathcal{L}_{\operatorname{EM}}+\mathcal{L}_{\operatorname{DAL-CRF}}$ & \textbf{94.16}	& \textbf{87.16}	& \textbf{91.83} &	87.42	& \textbf{90.14}
\\ \hline
\rowcolor{gray!5} DenseCRF~\citep{krahenbuhl2011efficient}-Postprocessing &  82.70	& 79.39	& 80.90	& 73.76 & 79.18 
  \\ 
\hline
\end{tabular}
}
\caption{Ablation Study on Different Components of the OD segmentation with loose prompts, in terms of Dice score. }
\label{tab:ablation}
\vspace{-0.6cm}
\end{table}
\textbf{Experiment Settings.} \textbf{Appearance Perturbations }: Note that our adopted multi-center datasets have strong domain shifts with appearance variations, we thus ignore test-time appearance perturbations (\textit{e.g.,} changes in contrast, noise, and blur). \textbf{Prompt Perturbations}: We mainly consider three kinds of bounding box prompts to evaluate TTA performance.  (i) \textit{Perfectness prompt (PP)}: The perfect prompt can be obtained by ground-truth (GT) mask, where the top-left and lower-right coordinates of the PP correspond to the minimal and maximal horizontal and vertical coordinates of the GT mask, respectively. A unique PP is utilized for TTA. (ii) \textit{Looseness prompts (LP)}: Clinically, it is labor-intensive and time-consuming to obtain the PP. Instead, a moderately loose prompt can be efficient but may trigger the degradation of the model with the need for TTA. Therefore, we define the looseness rate $\sigma$, which is the ratio of the coordinate shift relative to the width $w$ and height $h$ of the PP, to simulate real-world loose scenarios as follows:
\begin{equation}
    x(y)_{tl}^{LP}: x(y)_{tl} - \sigma \cdot w(h), x(y)_{lr}^{LP} : x(y)_{lr} + \sigma \cdot w(h), \nonumber
\end{equation}where $x(y)_{tl}$ and $x(y)_{lr}$ denote the vertical (horizontal) coordinates of the top-left and lower-right points of the PP. Each loose prompt is represented as $(x_{tl}^{LP},y_{tl}^{LP},x_{lr}^{LP},y_{lr}^{LP})$. Here, $\sigma$ is set to 0.1 and 0.2, as we observe 1) too small $\sigma$($<0.1$) is close to the PP without obvious changes and 2) too large $\sigma$($>0.2$) will usually result in significantly overlaps with other RoIs with unacceptable initial results. We report the average of different loose rates for comprehensive evaluation. (iii) \textit{Shrinkage prompts (SP)}: Similarly, we define the shrinkage ratio $\gamma$ to simulate the scenario that the bounding box prompt shifts to the interior of the PP as follows:
    \begin{equation}
    x(y)_{tl}^{SP}: x(y)_{tl} + \gamma \cdot w(h), x(y)_{lr}^{SP} : x(y)_{lr} - \gamma \cdot w(h). \nonumber
\end{equation}
Each shrinkage  prompt is represented as $(x_{tl}^{SP},y_{tl}^{SP},x_{lr}^{SP},y_{lr}^{SP})$. Similarly, $\gamma$ is set to 0.1 and 0.2. We report the average of different shrinkage rates for comprehensive evaluation.

\textbf{Baselines.} In this study, two kinds of TTA approaches for medical image segmentation are utilized for comparison. (1) \textbf{Normalization-based TTA}: Most baseline methods conduct the update of affine-transformation parameters and manipulation of source-trained statistics in BN layers, which the MedSAM lacks due to the usage of LN layers. We, therefore, discard source-trained statistics-related components and only optimize the affine-transformation parameters for baseline methods, including \textbf{TENT} (entropy minimization), \textbf{InTEnt} (foreground-background-balanced entropy weighting), \textbf{GraTa} (augmentation-based gradient alignment), and \textbf{PASS} (input decorator and alternating momentum updating strategy). (2) \textbf{Full update-based TTA}: The whole model parameters are updated based on customized losses, including \textbf{MEMO} (augmentation-based marginal entropy minimization) and \textbf{CRF-SOD} (sparse CRF regularization loss).  We leverage their open-source codes to apply into the architecture of MedSAM. To be fair, all baseline methods adopt a one-step optimization strategy. 

\textbf{Quantitive Analysis.} From Tables \ref{tab:comparisons_fundus_OC}, \ref{tab:comparisons_fundus_OD}, and \ref{tab:comparisons_spin}, some conclusions can be summarized. (1) Our proposed method without parametric updates surpasses all baselines in most cases with the best or second-best performance, showcasing the effectiveness of latent refinement. (2) It seems that full parameter updates are more suitable to large size RoIs than normalization-based updates, \textit{e.g.,} the OD. (3) For very challenging OC and SCGM tasks, most parametric update-based approaches have compromised performance compared with direct inference, but our proposed method performs well.

\textbf{Qualitative Analysis.} From Figure \ref{visualization latent},  our proposed method can achieve gradually better TTA segmentation with iterations (Note: we reduce the learning rate with multistep optimization rather than a one-step manner for understanding the optimization process better). Meanwhile, the latent embedding is gradually unsmoothed, which may be reasonable as the DAL-CRF loss enforces more statistics over the input image, leading to more discriminative information. The gradual decrease of DAL-CRF loss in each spatial point of the latent embedding is also observed, showcasing its effectiveness.

\textbf{Computation Complexity.} As illustrated in Table \ref{gflops}, we achieve around \textbf{7} $\times$ GFLOPs reduction compared with common usages. This is reasonable as 1) sophisticated augmentations (\textit{e.g.,} MEMO) and multiple gradient backward processes (\textit{e.g.,} GraTa) are computationally intensive. 2) For normalization-based updates, the gradient computation over normalization layers still passes through the entire model during backpropagation, which is unfavorable as MedSAM's image encoder is sophisticated with considerable consumption. 3) We circumvent any augmentation and the forward/backward process to the image encoder in a more efficient manner.

\textbf{Ablation Study \& Other TTA losses under the latent refinement framework.} As shown in Table \ref{tab:ablation}, $\mathcal{L}_{\operatorname{DAL-CRF}}$ significantly contributes more compared to entropy minimization loss and simply post-processing predictions fail to achieve acceptable performance on challenging medical images with strong appearance perturbations. Meanwhile, $\mathcal{L}_{\operatorname{DAL-CRF}}$ still performs better, even though other TTA losses are integrated into the latent refinement framework as shown in \hyperref[tab:TRAIN parameter]{5(b)}. Compared to the CRF-SOD (using input and output spaces), our DAL-CRF exhibits superiority using the latent-based co-occurrence regularization for foundation models.
\vspace{-0.3cm}
\section{Conclusion} We propose a novel TTA framework for foundation medical segmentation without parametric updates. Our theoretical analysis shows that direct latent refinement is feasible within the MedSAM architecture. By maximizing factorized conditional probabilities through our DAL-CRF loss combined with entropy minimization, we achieve better performance in terms of effectiveness and efficiency.
{
    \small
    \bibliographystyle{ieeenat_fullname}

}

\end{document}